\documentclass[runningheads]{llncs}
\usepackage{graphicx}
\usepackage{hyperref}

\usepackage{url}
\usepackage{color}

\begin{document}
\title{Security Architecture for Trustworthy Systems in 5G Era}
%
%
\author{Takayuki Sasaki\inst{1} \and
Shuichi Karino\inst{1} \and
Mikiya Tani\inst{1} \and
Kazuaki Nakajima\inst{1} \and
Koki Tomita\inst{1} \and
Norio Yamagaki\inst{1}
}
\authorrunning{T. Sasaki et al.}
%
\institute{
NEC, Japan\\
\email{\{tsasaki, karino, m-tani, nakajima\_k, koki-tomita, yamagaki\}@nec.com}
}
\maketitle              
\begin{abstract}
Systems using 5G are expected to be used in various cases of Society 5.0 and Industrie 4.0 such as smart cities, smart factories, and also critical infrastructures. These systems are essential for our life, thus cyberattacks against the system must be prevented. In this paper, we tackle two problems posed by 5G features: system construction using multi-vendor devices and softwarized functions. Specifically, there are supply-chain risks that malicious devices are used in the construction phase. Moreover, the softwarized network functions are easy to be attacked compared to hardware. To cope with these problems, we propose a concept of architecture comprising a blockchain to record security events including supply-chain information and a tamper detection engine to ensure the integrity of software components in 5G system. We implement the initial prototype of the architecture and show its feasibility.

\keywords{5G \and Security \and Tamper detection \and Blockchain}
\end{abstract}

\section{Introduction}
5G is a next-generation network architecture and it is expected to be used for various cases such as smart cities, smart factories, and operations of critical infrastructures. To use 5G for such cases, the security of the 5G system must be ensured because cyberattacks against such infrastructure cause serious damages to our society.

Security properties of 5G are improved from 4G.
For example, a security measure against IMSI~(International Mobile Subscriber Identity) catcher, which leaks user location information, has been introduced in 5G specifications.  However, we can identify the remaining security risks related to 5G features.
Specifically, we can identify two 5G features related to security. The first feature is the construction of 5G systems using multi-vendor devices. As 5G system is sophisticated and complex compared to 4G, we cannot construct a system with devices from a single vendor. Moreover, standard APIs are discussed and specified by 3GPP and ORAN alliance. In such a scenario, we need to take care of supply-chain risks that malicious devices are used in a system construction using third-party devices connected with the standard APIs.
The second feature is the softwarization of network functions. Specifically, Network Function Virtualization~(NFV) and Multi-access Edge Computing\footnote{It also called Mobile Edge Computing}~(MEC) are used in 5G systems. Moreover, in some cases of private 5G, 5G core components will be deployed on Cloud~\cite{virtual-core}. Attacks against such software are easy compared to attacks against the hardware due to the major vulnerabilities such as buffer overflow, modification of executable, and so on.
We can identify these features in 4G systems, but they are more obvious in 5G systems.

To cope with the above problems, we propose a concept of architecture with a blockchain to securely store supply-chain information and a tamper detection engine to measure the integrity of the software. The system information such as supply-chain information, tamper detection results, and system configurations are recorded into the blockchain. Using the data in the blockchain, an audit is performed to prove that the 5G system is secure.
Using the proposed architecture, we believe that the 5G systems can be used for critical infrastructure which requires strong security properties.

In summary, our major contributions are as follows:
\begin{itemize}
    \item A concept of secure 5G architecture to ensure the integrity of the entire 5G stack
    \item A blockchain-based event record mechanism for securely storing information including supply-chain information, tamper detection results, and system configurations used for a security audit of the 5G system.
\end{itemize}
\section{Problem statement}

\subsection{Supply-chain risks}
In 3G and 4G era, a system is constructed using a single vendor or a few vendors. However, in the 5G era, each system tends to be constructed using multi-vendor devices. This is because the whole system and each device used in the system become complex and it is difficult for a single vendor to supply devices of the entire stack of the 5G system. Moreover, telecommunication carriers would prefer an open system to avoid enclosure by a single vendor. For example, the open interface of Radio Access Network~(RAN) is discussed and standardized in ORAN alliance~\footnote{ORAN alliance, \url{https://www.o-ran.org/}}.

Because of the above situation, we have to take care of supply-chain risks. In the case of construction using a trusted single vendor, the supply-chain risk is not so critical, because the vendor can control the supply-chain and enforces security-by-design policies for all developers.
However, in the case of construction using multi-vendor devices, we have to ensure the trustworthiness of all devices from the external suppliers. Specifically, it should be proved that all devices are constructed in a secure development manner, e.g. only using valid hardware and software components.

We solve this problem by recording supply-chain information into a blockchain in Section~\ref{sec:event-store}.

\subsection{Attacks against softwarized network functions in 5G}
Softwarization is a trend of network systems. Traditionally, network functions are realized by hardware, but recently many functions are implemented using software for flexibility and cost efficiency. For example, NFV realizes the flexible deployment of services by chaining small software components. In most cases, virtualized network functions are deployed on virtual machines. Moreover, for the quick and low-cost deployment of 5G systems, 5G core functions could be deployed on cloud services such as Amazon EC2.
In addition, in 5G systems, the MEC is introduced for minimizing latency by deploying a computing node near to the end-user location in the 5G core network~\cite{MEC}.

Considering the above situation, we can identify risks about attacks against such software. Specifically, in case that software of a device in 5G has a vulnerability, the attacker can exploit the vulnerability and compromise the device. Of course, software components are used in the 4G systems, but considering the above situation, the risk would be increased from the 4G system era.

We solve this problem by deploying tamper detection engines into devices of the entire 5G system in Section~\ref{sec:tamper-detection}
\section{Architecture}
\label{sec:architecture}
Here we propose architecture to cope with the problems discussed in the previous section.

\subsection{Overview}
\begin{figure}[t]
  \begin{center}
    \includegraphics[width=12.0cm]{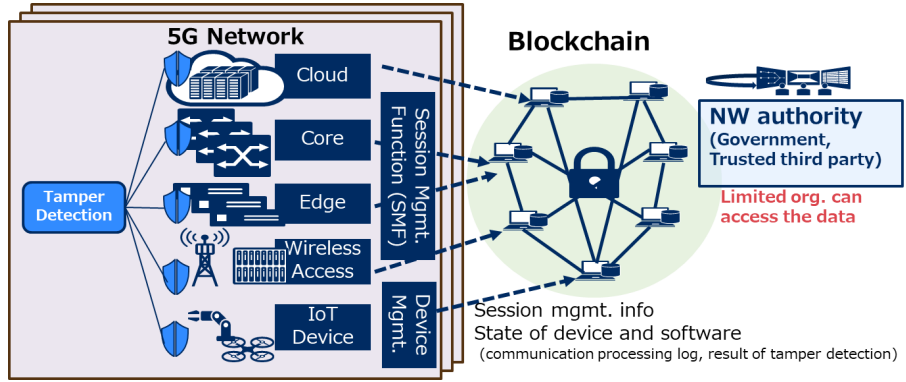}
    \caption{Conceptual architecture}
    \label{fig:architecture}
  \end{center}
\end{figure}

The proposed architecture comprises user devices~(e.g. smartphones, IoT devices), 5G devices~(e.g. RAN and 5G core), servers~(Figure~\ref{fig:architecture}), a secure event storage using blockchain, and tamper detection engines.

To securely store the security events, a blockchain is deployed. Specifically, supply-chain information and other security-related information such as logs of security appliances are recorded into the blockchain. Using this information, an auditor proves that the system is in the clean state and correctly operated.

To ensure the trustworthiness of the entire system, all layers should be protected. To this end, tamper detection engines are deployed in each device of all layers. The detection results are sent to the blockchain and used for security audits as well as the other information. 

\subsection{Secure event storage using blockchain}
\label{sec:event-store}
Blockchain is a distributed and trusted database that ensures the integrity of the recorded data. Specifically, each block of the blockchain contains a hash value of the previous block, thus malicious modification of a block requires modification of all hash values prior to the block. To ensure the consistency of the data, blockchain uses a consensus algorithm such as Proof of Work~(PoW) and Proof of Stake~(PoS).

There are two types of membership of blockchain: permissionless and permissioned. As for permissionless blockchain, participation is open to everyone, for example in bitcoin case, everyone can mine the coins. As for permissioned blockchain, its membership is pre-defined and members with permission can only access the blockchain.
In our use case, the information is shared with limited device manufacturers and 5G system operators, thus permissioned blockchain is better.

In the general blockchain model, data recorded by the blockchain are shared with the members. To limit the members accessing the specific data, we use a blockchain supporting access control~\cite{satellite-chain}. Specifically, a satellite chain constructed by specific nodes is created and data on the satellite chain can only be accessed by the node owners of the satellite chain.

\subsubsection{Recording supply-chain information}
To ensure that the 5G system is constructed using valid devices, supply chain information can be recorded into the blockchain. Specifically, network and IoT device manufacturers register supply-chain information of their product, for example, a list of hardware and software components and makers of the components. To build a 5G system without devices including invalid components, the SIer of the 5G system can check the component list on the blockchain. Moreover, the operator and users of the 5G system can check the component list as well.

The above information is used for the audit as evidence, thus the information must not be modified. In the case of malicious manufactures/insiders, the information should be managed in a distributed manner and its integrity should be ensured. To meet this requirement, we adopt blockchain. 

\subsubsection{Other information to be recorded}
\label{sec:recorded-info}
In addition to the supply-chain information, other types of information can be recorded. The following are possible types of information to be stored in the blockchain.

\paragraph{Maintainance histories}
Histories about maintenance such as update and repair can be recorded into the blockchain. Using this information, an audit about the maintenance can be performed.

\paragraph{Event logs from security appliances.}
The security event logs from security appliances can also be recorded in the blockchain and be used for a security audit of the system. The information also is useful for incident handling. For example, logs of a firewall, an intrusion detection system, and an authentication/authorization mechanism can be recorded.

\paragraph{Network configurations.}
To ensure the correct behavior of the entire network system, the correctness of the network configuration is required in addition to the integrity of the devices. To this end, routing information and network slice configurations can be recorded.

\paragraph{System events.}
To monitor system behavior, system events can be recorded.
For example, we can use 5G-related events such as registrations of user equipment, creations of sessions, errors from the network facilities.

\paragraph{Tamper detection results.}
To ensure the integrity of the entire 5G system, tamper detection results are also recorded. We will discuss it in the next subsection.

\subsection{Tamper detection engine}
\label{sec:tamper-detection}
To cope with the attacks against the software, the tamper detection engines monitor the integrity of devices. Specifically, to detect malware-infected devices and tampered devices by the attacker, the tamper detection engine periodically measures the integrity of the software running on the devices. Then, the tamper detection engine sends a measurement report to the blockchain.

Here, a security risk is attacks against the tamper detection engine itself. Specifically, an attacker may disable the tamper detection engine, then the attacker performs the second action for the attacker's primary purpose. To protect the tamper detection engine, the engine can be deployed in Trusted Execution Environment~(TEE) which creates an isolated software execution environment~\cite{safes}.

\subsection{Audit/Inspection using the data on the blockchain}
A security console shows the entire 5G security status by retrieving the data from the blockchain. Specifically,  using the data of the tamper detection results, a system administrator can checks whether the system is clean or not. In case that there are compromised devices, the administrator performs the incident response to the devices based on the data.
The other information discussed in Section~\ref{sec:recorded-info} is also displayed on this console.
\section{Implementation and evaluation}

We adopt our tamper detection engine using TrustZone of ARM Cortex-A~\cite{safes} and deploy the tamper detection engine to a small robotic arm. As for the blockchain, we use our original implementation~\cite{satellite-chain}.
The tamper detection engine and a blockchain node are connected via HTTP and REST interfaces.

\begin{figure}[t]
  \begin{center}
    \includegraphics[width=12.0cm]{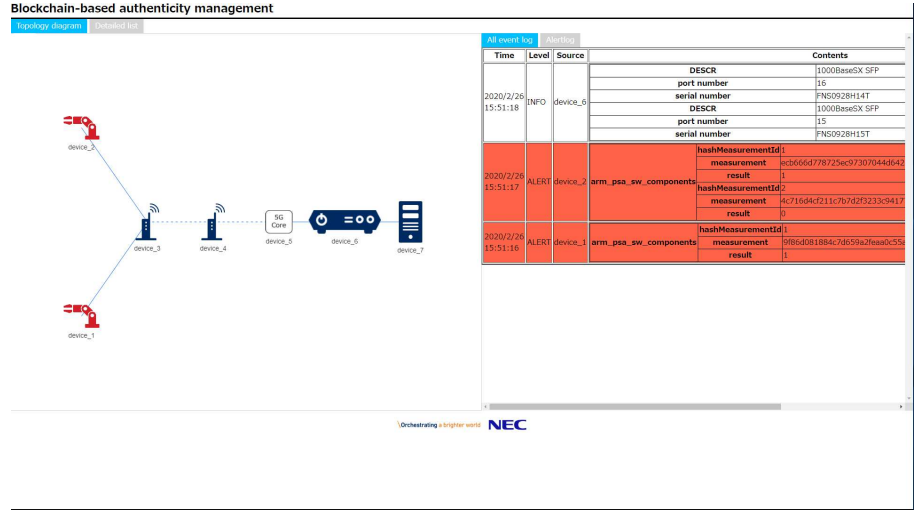}

    \caption{Prototype of a security management console}
    \label{fig:ui}
  \end{center}
\end{figure}

We also implement a security management console that shows the security status of the entire system~(Figure~\ref{fig:ui}). The console retrieves tamper detection results from the blockchain and shows the entire network topology and devices. In the figure, the compromised devices are shown using red icons.  On the right side of the figure, the table shows the details of the events including integrity measurement results of the robot arms. In addition, extra information such as supply-chain information recorded on the blockchain can also be displayed.

\section{Discussion}

\subsection{Integration with other security measures}
\label{other-security}
NIST cybersecurity framework\footnote{https://www.nist.gov/cyberframework} defines 5 functions~(identify, protect, detect, respond, recover) and the proposed architecture contributes to the identify function and the detect function.
Other security functions are also needed to be deployed. Specifically, protect, respond, and recover mechanisms should be integrated to the proposed architecture.
For example, major security functions such as authentication/authorization, intrusion prevention, DDoS mitigation should be deployed in addition to the proposed architecture. To integrate the proposed architecture and existing security functions, security logs from these security functions also can be recorded into the blockchain. And also, recover functions can leverage data in the blockchain to perform proper recovery from a compromised state.

\subsection{Information sharing across organizations}
For hardening the systems, security information needs to be shared across organizations such as the device manufacturers and users of the device~(e.g. 5G system operators). For example, vulnerability information and new attack methods information are useful for both of them.

However, some types of information such as specific network configurations are not willing to be shared because they reveal the internal structure of the network operators. In our architecture, accesses to such confidential information are controlled by the satellite chain\cite{satellite-chain} which can enforce access control.

\subsection{Who has blockchain nodes?}
Blockchain is a distributed database, thus there is a question; Who should have the blockchain nodes?
Currently, we assume that members of trusted consortium comprising device manufacturers, 5G system integrators have blockchain nodes. In addition, user organizations of 5G system can have the blockchain nodes. However, it depends on an operation model of the proposed architecture and we will further discuss this question as our future work.

\section{Related work}
Here, we introduce research about security architecture for 5G.
Rahimi et. al. have studied the security of 5G-IoT architecture and summarized attacks against the architecture~\cite{Rahimi}. Moreover, they have proposed 5G-IoT security taxonomy with five layers: an application layer, a communication layer, a network layer, a data layer, and a perception layer.
Our architecture does not cover all attacks pointed out by this work, thus our architecture needs to be integrated to existing security solutions as discussed in Section~\ref{other-security}

Blanc et al. have proposed security architecture relying on virtualization and softwarization~\cite{Blanc}. In this architecture, security functions are dynamically deployed in a security-as-a-service manner. Specifically, the architecture as two major functions: a security monitoring framework to detect security events and a security orchestrator reacts against the security events. The security services are provided as virtual functions and security service chaining is performed.
Adam and Ping have proposed a framework for security event management~\cite{adam}. The proposed framework performs automated event data collection across layers such as a telecommunication layer and an IaaS layer for monitoring of shared resources. The framework also performs automated mitigation of security SLA violations by reconfiguring security functions.
These architectures and our architecture are in the complementary relation. Specifically, the concept of tamper detection and the blockchain-based event storage can be integrated with their detection and response mechanisms.

Sinclair et. al have proposed an architecture for drug supply chain management using blockchain~\cite{drug-supply-chain}. Our architecture is an extension of their concept and uses system construction information and operation information.
\section{Conclusion and Future work}
To build a secure 5G system, secure monitoring and audit are essential, especially for critical infrastructures requiring strong security. To this end, we propose an architecture comprising the tamper detection engine and the secure event storage using blockchain. By inspecting the data on the blockchain, the security state of the 5G system can be proved.

As for future work, we will implement the entire system and conduct evaluation using a 5G system for pointing out deployment and operation issues. Moreover, we will consider non-technical matters such as trusted partnerships and collaborative operations among device manufacturers, system integrators, and 5G system operators.

This research is still a work in progress, but we believe that the proposed architecture boosts applications of 5G systems to critical infrastructures.

\bigskip \noindent\textbf{Acknowledgements.}
The authors would like to thank Ghassan Karame for feedback about the concept of the architecture. The authors also would like to Seng Pei Liew for reviewing the paper.

%
%
%
\bibliographystyle{splncs04}
\bibliography{bibliography}

\end{document}